# All-optical reconfigurable chiral meta-molecules

Linhan Lin[1,2,*], Sergey Lepeshov[3], Alex Krasnok[4,6], Taizhi Jiang[5], Xiaolei Peng[2], Brian A. Korgel[2,5], Andrea Alù[4,6,7,8] and Yuebing Zheng[1,2,*]

[1]Department of Mechanical Engineering, The University of Texas at Austin, Austin, TX 78712, USA

[2]Materials Science & Engineering Program and Texas Materials Institute, The University of Texas at Austin, Austin, TX 78712, USA

[3]ITMO University, St. Petersburg 197101, Russia

[4]Department of Electrical and Computer Engineering, The University of Texas at Austin, Austin, TX 78712, USA

[5]Mc Ketta Department of Chemical Engineering, The University of Texas at Austin, Austin, TX 78712, USA

[6]Photonics Initiative, Advanced Science Research Center, City University of New York, New York, NY 10031, USA

[7]Physics Program, Graduate Center, City University of New York, NY 10016, USA

[8]Department of Electrical Engineering, City College of The City University of New York, NY 10031, USA

[*]Corresonding authors at: Department of Mechanical Engineering, The University of Texas at Austin, Austin, TX 78712, USA

*E-mail addresses*: Lin, L. (linlh04@utexas.edu), Zheng, Y. (zheng@austin.utexas.edu)

**Abstract**

Chirality is a ubiquitous phenomenon in the natural world. Many biomolecules without inversion symmetry such as amino acids and sugars are chiral molecules. Measuring and controlling molecular chirality at a high precision down to the atomic scale are highly desired in physics, chemistry, biology, and medicine, however, have remained challenging. Herein, we achieve all-optical reconfigurable chiral meta-molecules experimentally using metallic and dielectric colloidal particles as artificial atoms or building blocks to serve at least two purposes. One is that the on-demand meta-molecules with strongly enhanced optical chirality are well-suited as substrates for surface-enhanced chiroptical spectroscopy of chiral molecules and as active components in optofluidic and nanophotonic devices. The other is that the bottom-up-assembled colloidal meta-molecules provide microscopic models to better understand the origin of chirality in the actual atomic and molecular systems.

**Introduction**

Chirality, the property of an object that is non-superimposable to its mirror image, is quite ubiquitous in the living world. Many biomolecules, including amino acids, sugars, proteins, and DNA, are chiral. These chiral molecules and their mirror images (enantiomers) exhibit opposite handedness when interacting with circularly polarized light, as well as different chemical and biological performance. Inspired by chiral molecular structures, researchers are developing strategies to build artificial chiral materials by mimicking molecular structures using functional materials [1-14]. Specifically, metal nanomaterials exhibit tailorable optical properties upon excitation of surface plasmons and become one of the most promising components to realize of chiral optical metamaterials [15,16]. In the past decade, top-down fabrication of plasmonic meta-molecules has been extensively studied [17]. However, top-down fabrication requires sophisticated instruments and multiple-step process, and the fabrication of multiple-composition and reconfigurable chiral materials is still challenging.

Colloids have better similarity to atoms for the bottom-up construction of molecular structures at colloidal scale. The diversity of materials, sizes, and functionality of colloids provides abundant building blocks for the assembly of complex chiral colloidal molecules [18-20]. For example, using magnetic colloids or asymmetric colloidal dimers as building blocks, chiral colloidal molecules have been successfully synthesized *via* an external magnetic field [21] or electric field [22]. Atomic force microscopy has been used to manipulate and assemble metallic colloids into chiral structure [23], while it is not an efficient approach with extremely low throughput. DNA nanotechnology has proven a versatile tool to build complex colloidal molecules [24-28]. Specifically, DNA origami provides a programmable template for a precise organization of colloids into a variety of two-dimensional (2D) or three-dimensional (3D) chiral architectures [29-31]. Incorporation of photoresponsive or pH-sensitive DNA lockers onto the DNA origami allows dynamic control of the geometry and optical chirality through external stimuli [32]. However, the tunability is still limited to a few pre-established discrete states and on-demand reconfigurability is still elusive.

On-demand manipulation and organization of colloidal molecules using light is attractive in nanotechnology and nanophotonics. Optical tweezer is a versatile optical tool which can handle colloids in 3D manner [33]. However, the high optical power required in optical tweezers can easily damage the functional materials or molecules and alter their intrinsic physical or chemical properties. The optical trapping of optical nanomaterials, such as metallic or high-refractive-index dielectric colloids is even more challenging due to the strong radiation force arising from the enhanced scattering or absorption [34,35]. In addition, the bonding between colloids relies on external interaction beyond optical trapping force and all-optical assembly cannot be achieved. Herein, taking advantage of the light-generated thermoelectric field and the micelle-mediated colloidal bonding [36-38], we demonstrate the all-optical atom-by-atom assembly of chiral meta-molecules and *in-situ* measurement of the optical chirality. Specifically, chiral meta-molecules can be disassembled into isolated meta-atoms for re-organization into their enantiomers (or other isomers), which enables on-demand modulation of the optical chirality. Although the atomic interaction and chiral origin are different between organic chiral molecules and the artificial chiral meta-molecules, some common parameters such as geometric asymmetry and composition asymmetry exist in both systems. Therefore, the all-optical and reconfigurable assembly of chiral meta-molecules provides a simplified platform to understand chirality at the colloidal scale, which will open a new door for on-demand manufacturing of chiral metamaterials for biosensing [39-41] and active chiral photonic devices [42-46].

**Results and discussion**

*Concept and Working Principle*

Figure 1a shows the design concept of the chiral meta-molecules. Taking a simple close-packed colloidal trimer as an example, chirality does not exist in trimers consisting of three meta-atoms with the same size and material. However, when size asymmetry or composition asymmetry is imposed onto the structure, the trimer becomes chiral in the two-dimensional plane, i.e., planar-chiral. It should be noted that, when the spherical centers of all the meta-atoms can be placed in the same plane (e.g., all the chiral trimers), the chiral meta-molecules can be superimposed to their mirror images via three-dimensional rotation. However, upon analysis with the directional incident light, the meta-molecules exhibit chiral optical response. That is to say, the metamolecules "seem" to be chiral under the directional light incidence. In addition, the presence of substrate further breaks the symmetry of the metamolecules because of the addition of out-of-plane asymmetry. Such meta-molecules are termed planar-chiral meta-molecules in Fig. 1a. However, when more meta-atoms are added into the meta-molecules and the spherical centers of all the meta-atoms cannot be placed in the same plane, the meta-molecules become real-chiral even without considering the effect from the substrate or the directional incident light. Although the design concept is theoretically simple, it is technically difficult to select and position the meta-atoms precisely. To address this challenge, we use a light-generated temperature field to drive the ionic separation in the fluid, which further creates a thermoelectric field to trap the colloidal particles at the laser spot [36]. Individual meta-atoms are delivered and assembled into specific molecular structures by steering the laser beam (Fig. 1b) [38]. Dark-field scattering spectra are recorded *in-situ* under different circularly polarized light excitation when the meta-molecules are held in the opto-thermoelectric field (see supplementary Fig. 1 for the optical setup). The meta-molecules are dis-assembled into isolated meta-atoms when the laser is turned off and the meta-atoms can be re-assembled into their enantiomers with different handedness on-demand.

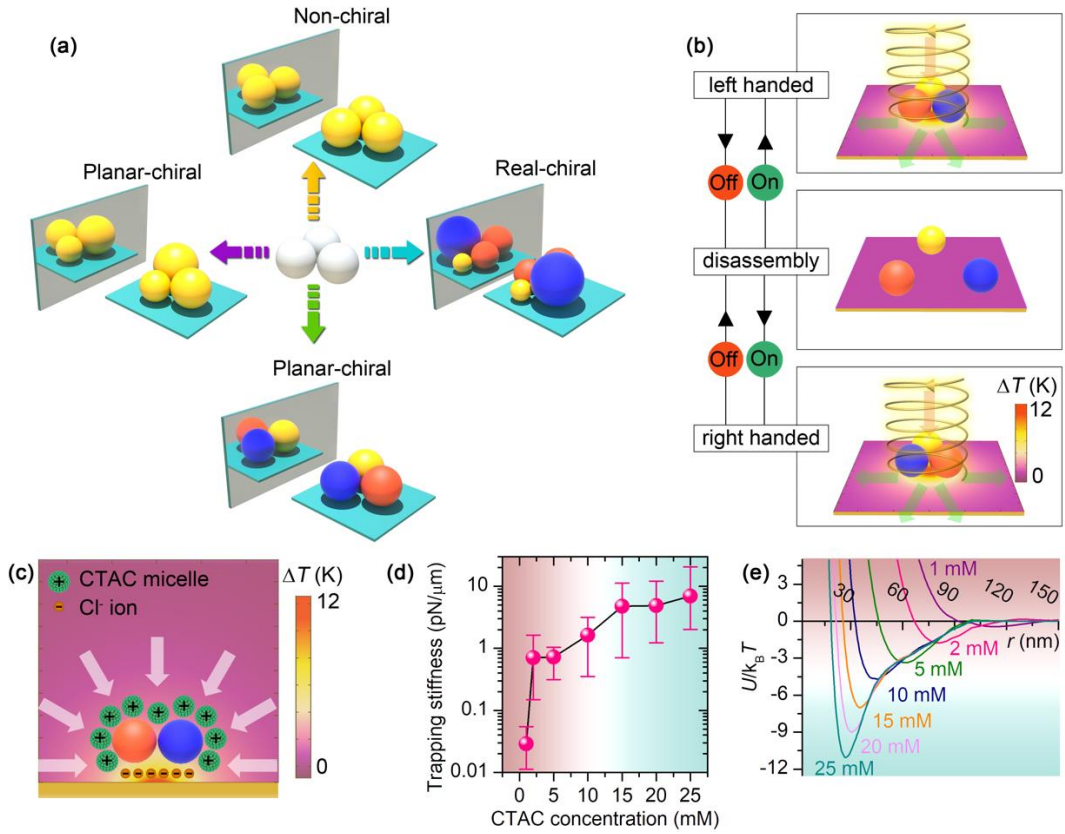

**FIGURE 1.** Concept and working principle of reconfigurable chiral meta-molecules. (a) Chirality origin in colloidal meta-molecules. The balls with different colors represent colloids with different materials. (b) Working principle of the reconfigurable opto-thermophoretic assembly of chiral meta-molecules. The red and green arrows represent the incident and scattering light, respectively. (c) Colloid-micelle interaction in the light-generated temperature field. (d) Measured trapping stiffness of single 300 nm AuNSs measured at different CTAC concentrations. (e) Inter-particle interaction potential $U$ between two 300 nm AuNSs at different CTAC concentrations. The green color indicates stable trapping or assembly region, while red color indicates that the trapping or assembly is not stable. $T$ is the temperature and $k_B$ is the Boltzmann constant.

It should be noted that the optical chirality is highly sensitive to the structural configuration of the chiral meta-molecules, and the stability of the meta-atoms in the optical trap is the key for the optical characterization. Fig. 1c shows the underlying trapping mechanism of the assembly strategy, where the thermoelectric field arising from the spatial separation between the cetyltrimethylammonium chloride (CTAC) micelles and the Cl$^-$ counterions and the depletion attraction force from the thermophoretic migration of the CTAC micelles provide the major trapping force [36]. The structural stability of the meta-molecules in the optical trap relies on both the micelle-meta-atom interaction and the meta-atom-meta-atom interaction, i.e., the trapping stiffness and the interparticle interaction potential. Taking 300 nm gold nanoparticles (AuNPs) as an example, we measured the trapping stiffness of single 300 nm AuNPs at different CTAC concentrations (Fig. 1d and Supplementary Fig. 2). Above 15 mM, the trapping stiffness on a single particle reaches more than 4 pN/μm, which represents a stable trapping at the laser spot. Moreover, we calculated the interaction potential between two 300 nm AuNPs in the optical trap, as shown in Fig. 1e (see Supplementary Note). We can see that the potential well is below -6 $k_BT$ at 15 mM, which can be further optimized by increasing the CTAC concentration. It should be noted that the interaction potential is a function of the particle size, which requires a higher CTAC concentration for the assembly of smaller meta-atoms [38]. The typical optical power used for the assembly is about 0.15 mW, which is 2-3 orders of magnitude lower than that in optical tweezers. The low optical power significantly reduces the thermal perturbation caused by the laser beam. More importantly, at such a low optical power, there is no radiation damage to the colloids.

*Plasmonic Chiral Trimers and Dielectric Chiral Trimers*

The capability of manipulating single colloidal particles with different materials and sizes using the light-directed thermoelectric field provides a general strategy to assemble a variety of chiral meta-molecules. Here, we use both plasmonic nanoparticles and hydrogen-terminated amorphous silicon nanoparticles (SiNPs) with high refractive index [47] as the building blocks. The dielectric function of the SiNPs is tunable by controlling the hydrogen concentration inside the particles. In this work, SiNPs with two different hydrogen concentrations, i.e., 5 at.% and 20 at.% (noted as 5H and 20H), were synthesized for optical assembly. Figs. 2a-c show both the experimental and simulated dark-field scattering spectra from individual 300 nm AuNP, 300 nm SiNP (5H), and 500 nm SiNP (5H) which were optically hold in the opto-thermoelectric field, respectively. Specifically, the excitation of multiple high-quality dielectric resonances inside the SiNPs enriches the spectral features of the building blocks for optical design of the chiral meta-molecules [48]. The single-particle scattering spectra of SiNPs were calculated using Mie theory based on the permittivity of amorphous Si (see Methods). The slight mismatching between the experimental and theoretical spectra, in particular, the disappearance of the high-order multipole modes ($l$=4, $l$=5) of the 500 nm SiNP, can be explained by a slightly elliptical shape of the actual particles. Moreover, the SiNPs used here also contain Si nanocrystals in the particles, which introduce a peak at 503 cm$^{-1}$ in the Raman spectrum (Supplementary Fig. 3d). The co-existence of both amorphous silicon and nanocrystalline silicon can introduce an offset in the permittivity and thus cause the minor mismatching between the experimental and simulated spectra.

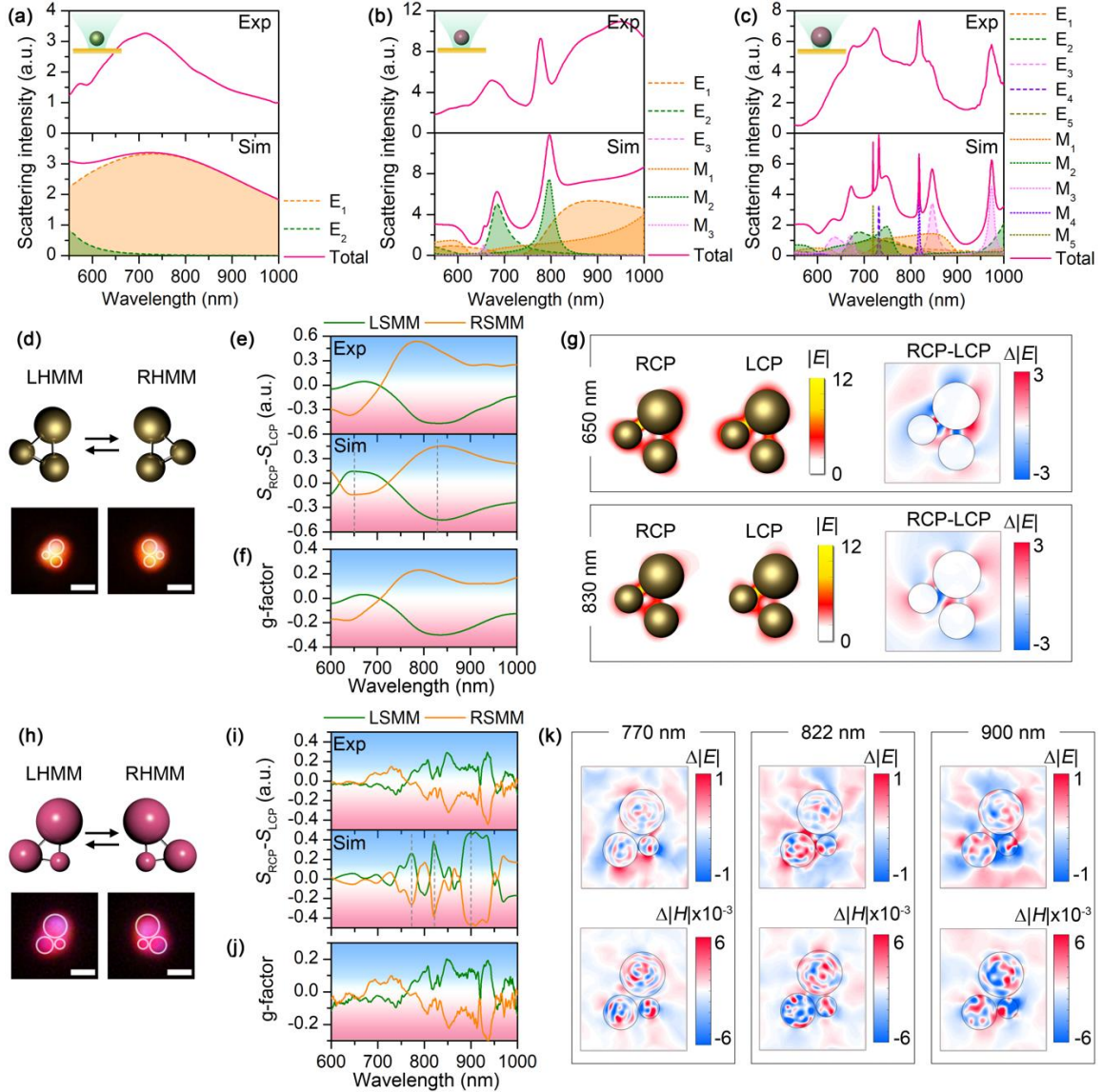

**FIGURE 2.** Building blocks of the chiral meta-molecules, and the Au and Si chiral trimers. (a-c) Experimental (top) and simulated (bottom) scattering spectra of single 260 nm AuNP (a), 300 nm SiNP (5H) (b), and 500 nm SiNP (5H) (c). Multipole decomposition is carried out for the simulated spectra, with the electrical ($E_l$) and magnetic ($M_l$) spherical Mie modes of $l$ orders, i.e., $l$=1: dipole mode; $l$=2: quadrupole mode; $l$=3: octupole mode; $l$=4: hexadecapole mode, etc. (d) Schematic and optical images of left-handed and right-handed Au chiral trimer. (e) Experimental (top) and simulated (bottom) differential scattering spectra of the Au chiral trimer. (f) G-factor of the Au chiral trimer. (g) Electric field distribution and differential electric field distribution of the Au chiral

trimer at the wavelengths of 650 nm (top) and 830 nm (bottom). (h) Schematic and optical images of left-handed and right-handed Si chiral trimer. (i) Experimental (top) and simulated (bottom) differential scattering spectra of the Si chiral trimer. (j) G-factor of the Si chiral trimer. (k) Differential electric and magnetic field distribution profiles of the Si chiral trimer at 770 nm (left), 822 nm (middle) and 900 nm (right). In (d-g), the diameters of the AuNPs are 230 nm, 260 nm, and 330 nm, respectively. In (h-k), the diameters of the SiNPs are 300 nm, 500 nm, and 700 nm, respectively. In (d) and (h), white circles are superimposed onto the dark-field optical images to guide the visualization of the molecular structures. Scale bars: 1 µm.

As a first demonstration, we assembled an Au chiral trimer using AuNPs of different sizes. Both the left-handed meta-molecule (LHMM) and right-handed meta-molecule (RHMM) were assembled in the opto-thermoelectric field with the same three AuNPs through steering the light field, with both the schematic and optical images displayed in Fig. 2d. The color in the dark-field optical images arises from the wavelength-dependent scattering of the particles. To characterize the optical chirality, we measured the forward scattering spectra of the chiral meta-molecules with incident light at both right-handed circular polarization (RCP) and left-handed circular polarization (LCP), and calculate the differential scattering spectra by $S_{\text{RCP}} - S_{\text{LCP}}$, where $S_{\text{RCP}}$ and $S_{\text{LCP}}$ are the scattering spectra at right-handed and left-handed polarization, respectively (see Methods and Supplementary Fig. 1 for the optical setup). The scattering spectra were normalized by the incident light, e.g., $S_{\text{RCP}} = \frac{I_{\text{RCP}}^{\text{scat}}}{I_{\text{RCP}}^{\text{in}}}$, where $I_{\text{RCP}}^{\text{scat}}$ and $I_{\text{RCP}}^{\text{in}}$ are the measured intensity of the scattering light and incident light, respectively. It should be noted that the detection is rotationally symmetric in the two-dimensional (2D) plane and the 2D rotation on the substrate will not change the measurement results, which simplifies the assembly process of the chiral meta-molecules. We can see that the LHMM shows a peak at 650 nm and a dip at 830 nm in the differential scattering spectrum, respectively. The differential scattering spectra of both LHMM and RHMM present a bisignate feature, indicating mirror symmetry between the enantiomers. It should be noted that the mirror symmetry is not perfect in the Au trimer due to the imperfect spherical shape of the AuNPs, which induces minor geometric anisotropy from single nanoparticles. We calculate the asymmetric factor (g-factor) of the meta-molecules by $g = \frac{2(S_{\text{RCP}} - S_{\text{LCP}})}{S_{\text{RCP}} + S_{\text{LCP}}}$, as shown in Fig. 2f. A high g-factor of 0.2 is obtained, revealing excellent chiral response from the Au chiral trimer. To understand the origin of chirality from the Au chiral trimers, we simulated the LHMM numerically, which shows a reasonable spectral matching with the experimental results (bottom panel of Fig. 2e). The electric field distributions under irradiation at two different circular polarizations are illustrated in Fig. 2g. Specifically, the distinctive feature can be clearly seen in the differential electric field distribution in the right panel, showing that the interparticle coupling between the 330 nm AuNP and the two smaller AuNPs gives rise to the chirality at both frequencies, while the direct coupling between the 230 nm AuNP and the 260 nm AuNP has minor contribution.

The optical Mie resonances from the SiNPs provide a rich fingerprint to tailor the optical chirality. As a demonstration, we assemble a chiral trimer using SiNPs of different sizes, as shown in Fig. 2h. The excellent mirror symmetry of the differential scattering spectra between the left-handed and right-handed Si trimers (Fig. 2i) made of the same metaatoms represents a precise control of the molecular structure. Compared with the AuNPs, the almost spherical shape of the SiNPs (Supplementary Fig. 3) gives a better matching between the experimental geometry and the theoretical design. From the original scattering spectra of the Si chiral trimer (Supplementary Fig. 4), we can see a maximum intensity variation of ~35% between two different circular polarizations, suggesting an excellent chiral effect in the Si chiral trimer. Specifically, both the experimental and simulated differential scattering spectra present multiple chiral modes with high quality factors, which arises from the high-quality optical Mie resonances in individual SiNPs (see Supplementary Fig. 5). Interestingly, the *in-situ* spectroscopic characterization during the assembly process makes it possible to compare the differential scattering spectra with the scattering spectra from individual particles, which shows a good frequency matching (Supplementary Fig. 5) for some chiral features and indicates that the interparticle coupling under different circular polarizations significantly modifies the scattering intensity of individual dielectric resonances to generate optical chirality. The wavelength-dependent g-factor (Fig. 2j) shows a similar feature with the differential scattering spectra, with a maximum value of 0.2-0.3 observed at 940 nm. To further investigate the role of electromagnetic coupling in the chirality, we select three chiral modes at 770 nm, 822 nm, and 900 nm, and calculated the differential electric field and magnetic field distribution, respectively (also see Supplementary Fig. 6 for the raw field distribution). From Fig. 2k, we can see that the RCP light strengthens both the 500 nm SiNP-300 nm SiNP and 500 nm SiNP-700 nm SiNP couplings at 822 nm, while it weakens the 300 nm SiNP-500 nm SiNP at 770 nm, and both the 300 nm SiNP-500 nm SiNP and 300 nm SiNP-700 nm SiNP couplings at 900 nm. Bridging between the chiral response and the interparticle coupling opens a door to understand and tailor the optical chirality rationally.

*Versatility of the Chiral Meta-Molecules*

Beyond the Au and Si chiral trimers, the versatility of the assembly strategy allows us to build diverse chiral meta-molecules using a wide range of meta-atoms with different sizes and materials. In Figs. 3a and b, we build chiral trimers using meta-atoms with the same nominal sizes but different materials. In term of geometric symmetry, the meta-molecules are expected to be achiral. However, the composition asymmetry in the meta-molecules breaks the inversion symmetry. As shown in Fig. 3a, the assembly of an AuNP, an AgNP, and a SiNP (5H) with the same diameter of 300 nm gives rise to a distinct chiral mode at 730 nm, which is consistent to the magnetic quadrupole ($M_2$) mode from the single 300 nm SiNP (5H). The incorporation of SiNPs with different hydrogen contents can also induce the planar chirality. The meta-molecule composed of a 300 nm AuNP and two 300 nm SiNPs with different hydrogen contents shows a multiband chiral response, as shown in Fig. 3b. We further demonstrate the chiral trimers by incorporating both geometric and compositional asymmetry into the single chiral meta-molecules (Figs. 3c-d and Supplementary Fig. 7).

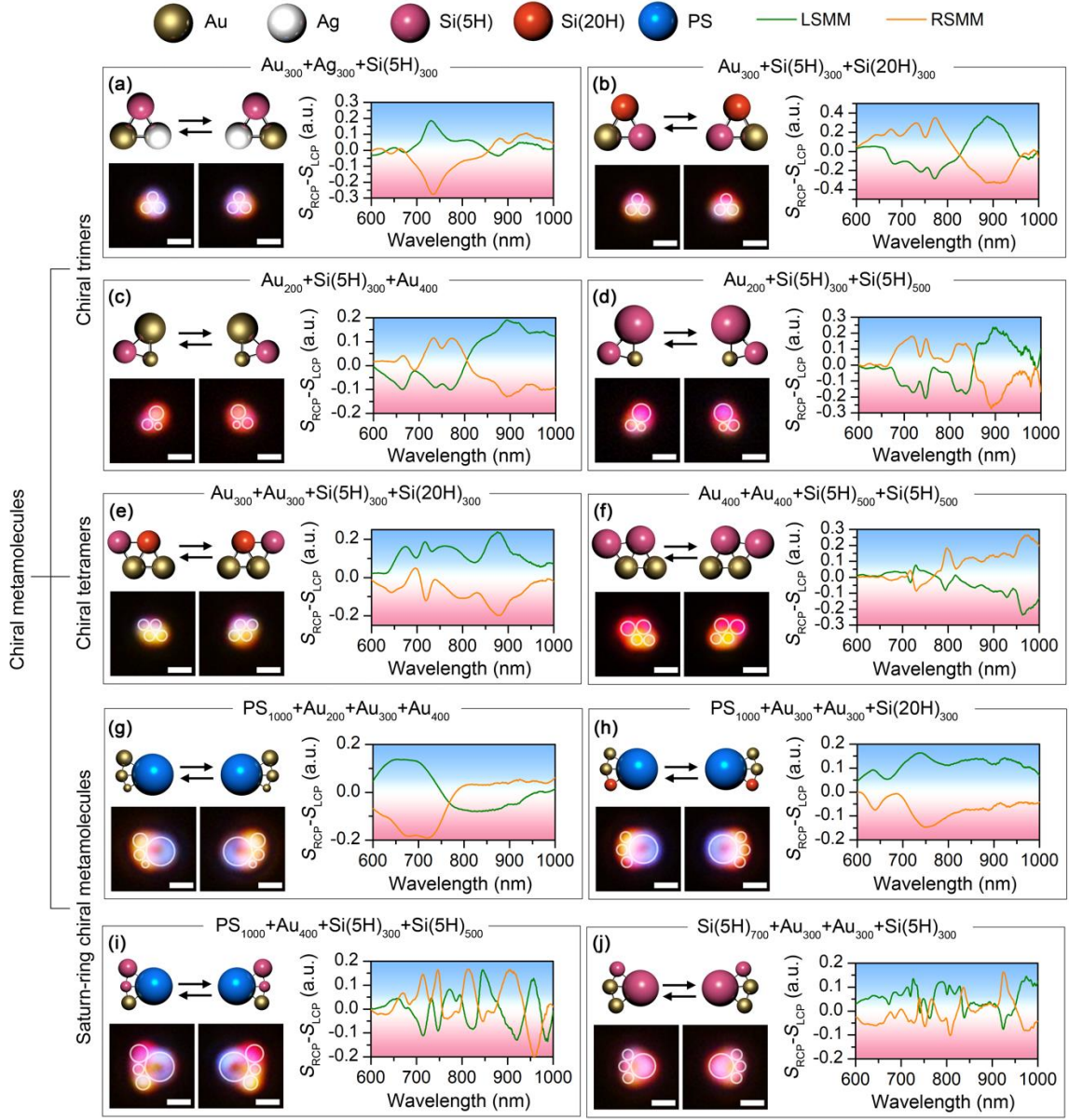

**FIGURE 3.** Versatility of the experimentally assembled chiral meta-molecules. Schematic, optical images and differential scattering spectra of chiral trimers composed of (a) an AuNP, an AgNP and a SiNP (5H) with diameter of 300 nm; (b) an AuNP, a SiNP (5H) and a SiNP (20H) with diameter of 300 nm; (c) a 200 nm AuNP, a 300 nm SiNP (5H), and a 400 nm AuNP; (d) a 200 nm AuNP, a 300 nm SiNP (5H) and a 500 nm SiNP (5H). Schematic, optical images and differential scattering spectra of chiral tetramers composed of (e) two 300 nm AuNP, a 300 nm SiNP (5H), and a 300 nm SiNP (20H); (f) two 400 nm AuNPs and two 500 nm SiNPs (5H). Schematic, optical images and differential scattering spectra of Saturn-ring chiral meta-molecules composed of (g) a 1 µm PS bead, a 200 nm AuNP, a 300 nm AuNP, and a 400 nm AuNP; (h) a 1 µm PS bead, two 300 nm AuNPs, and a 300 nm SiNP (20H); (i) a 1 µm PS bead, a 400 nm AuNP, a 300 nm SiNP (5H), and a 500 nm SiNP (5H); (j) a 700 nm SiNP (5H), two 300 nm AuNPs, and a 300 nm SiNP (5H). The meta-molecules on the left are LHMMs, while the ones on the right are RHMMs. White circles are superimposed onto the dark-field optical images to guide the visualization of the molecular structures. Scale bar: 1 µm.

Building chiral meta-molecules with more meta-atoms and more complicated geometry is possible. Figs. 3e and f show a chiral tetramer built with two 300 nm AuNPs, one 300 nm SiNP (5H), and one 300 nm SiNP (20H) and a chiral tetramer built with two 400 nm AuNPs and two 500 nm SiNPs (20H), respectively (also see Supplementary Fig. 8 for other chiral tetramers). Taking a micron-sized polystyrene (PS) bead as a central meta-atom which provides an arc to arrange other meta-atoms, we demonstrate the Saturn-ring chiral meta-molecules (Figs. 3g-i). Specifically, the Saturn-ring meta-molecules composed of PS beads and AuNPs shows a broad chiral model at 700 nm (Fig. 3g). The incorporation of SiNPs into the chiral meta-molecules enriches the spectral feature with multiband and high-quality-factor chiral modes (Figs. 3h and i). We define quality-factor as $Q = \frac{\lambda_0}{\Delta\lambda}$, where $\lambda_0$ is the peak/dip wavelength of the chiral mode and $\Delta\lambda$ is the full width at half maximum of the chiral mode. For example, the RHMM in Fig. 3g gives a quality-factor of about 5 for the chiral mode at 700 nm. However, the RHMM in Fig. 3i gives a quality-factor of 35.9 for the chiral peak at 714 nm and 49.8 for the chiral peak at 747 nm. The use of a 700 nm SiNP as the central meta-atom is further demonstrated in Fig. 3j. Using a single laser beam, we have succeeded in the assembly of chiral meta-molecules with five meta-atoms considering the sizes of the temperature field and meta-molecules (Supplementary Fig. 9). Multiple laser beam is required to manipulate the meta-atoms in parallel to include more meta-atoms in the meta-molecules. To further verify the structural configuration of the chiral meta-molecules, we improved the inter-particle bonding as well as the substrate-particle bonding to print the meta-molecules on the substrate for scanning electron microscopy (SEM), as shown in Supplementary Fig. 10. The excellent matching between the morphology and the theoretical design further reveals the precise geometry control of the chiral meta-molecules. It is worth noting that all these chiral meta-molecules show high g-factors between 0.2 and 0.4 (Supplementary Fig. 11), which is among the highest values in the reported chiral structures [17,18,24,49-52].

To further reveal the relevance between the structure and the optical chirality of the meta-molecules, we demonstrate the step-by-step assembly process of a meta-molecule composed of a 1 μm PS, a 300 nm AuNP, a 300 nm SiNP (5H) and a 500 nm SiNP (5H) (see Supplementary video 1), with the optical chirality recorded at different assembly stages, as shown in Figs. 4a-d. No optical chirality is observed for the isolated PS bead and the 1 μm PS-500 nm SiNP (5H) dimer (Figs. 4a-b). The optical chirality shows up when the 300 nm AuNP (5H) is added to the dimer to obtain a chiral trimer (Fig. 4c). A new chiral feature appears at 788 nm when the 300 nm SiNP is further added into the meta-molecules (Fig. 4d). In the meantime, the sign is switched from positive to negative below 730 nm in the differential scattering spectra. The chiral-metamolecule was dis-assembled when the laser was turned off and the meta-atoms were re-assembled into the right-handed meta-molecule step-by-step, with the *in-situ* differential scattering spectra displayed in Figs. 4e-h. Similar spectral evolution was observed with mirror symmetry to that of the left-handed meta-molecule. It should be noted that the atom-by-atom assembly process provides an accurate control of the structural arrangement. All the meta-molecules have thermodynamically stable configurations with close-pack arrangement. Besides the optical images, the structural change of the meta-molecules can be monitored by the *in-situ* scattering spectra to make sure that stable molecular structures consistent with our schematics are obtained.

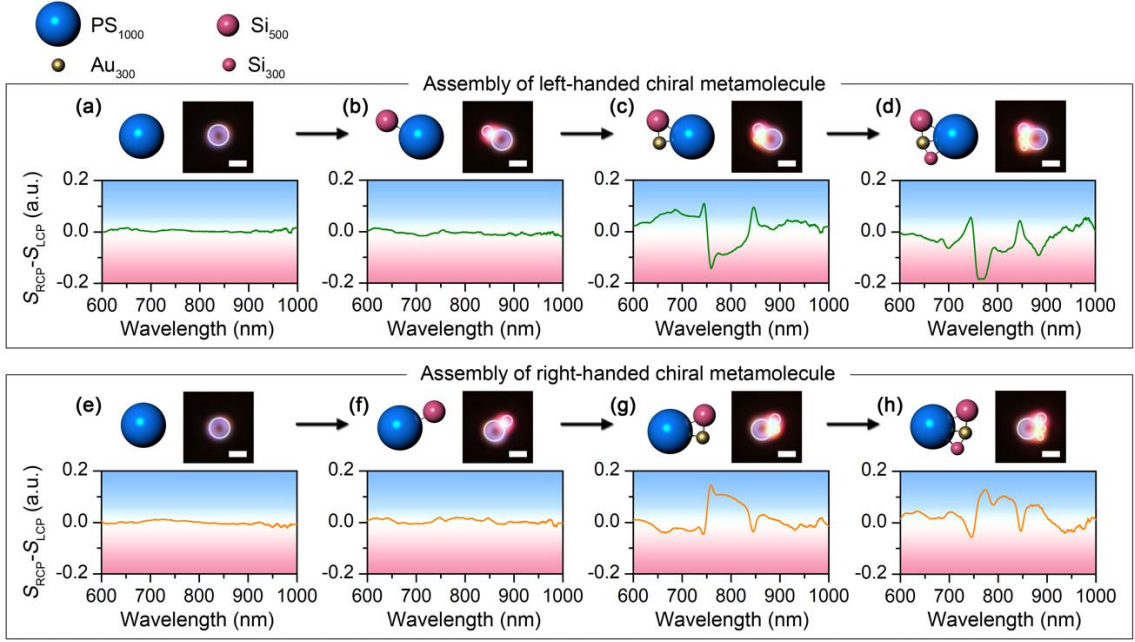

**FIGURE 4.** Assembly process of chiral meta-molecules of both handedness with *in-situ* measurement of the optical chirality. The schematics, optical images, and differential scattering spectra show the step-by-step assembly process of both the left-handed (a-d) and right-handed (e-h) chiral meta-molecules composed of a 1 μm PS, a 300 nm AuNP, a 300 nm SiNP (5H) and a 500 nm SiNP (5H). White circles are superimposed onto the dark-field optical images to guide the visualization of the molecular structures. Scale bar: 1 μm.

*On-Demand Configurational Control in Chiral Meta-Molecules*

It is well known that molecules show different handedness between their enantiomers. However, in other isomers, such as diastereomers, chirality control is still mysterious. For example, both D-threose and L-threose are the diastereomers of D-erythrose. However, D-erythrose and D-threose have the same handedness, while L-threose shows a different handedness (Supplementary Fig. 12). In other words, the switching of -H and -OH groups at the C2 chiral carbon center in D-erythrose does not change the handedness, while similar switching at C3 flips the handedness. In order to investigate the dependency of optical chirality on the structural configuration of the meta-molecules, we use a 1 µm PS bead, a 400 nm AuNP, a 500 nm SiNP (5H), and a 700 nm SiNP (5H) (Fig. 5a) to build three sets of chiral metamolecules and their corresponding enantiomers, as shown in Figs. 4b-d. The differential scattering spectra indicate that optical chirality is tunable by controlling the molecular geometry in the chiral meta-molecules. More interestingly, a direct comparison among the three meta-molecules shows the partial control of the handedness can be achieved by flipping the location of the meta-atoms. As shown in Fig. 5e, the chiral meta-molecule was built by switching the location between the 500 nm SiNP (5H) in the middle and the 400 nm AuNP at the bottom. The differential scattering spectra show mirror symmetry with respect to the zero line below 850 nm (the region superimposed with yellow color in Fig. 5e). It can be inferred that the 1 µm PS-400 nm AuNP-500 nm SiNP (5H) coupling dominates the chiral features below 850 nm, i.e., the switching of circular polarization between LCP and RCP significantly modifies the inter-particle coupling strength in the 1 µm PS-400 nm AuNP-500 nm SiNP (5H) trimer structure between 600 nm and 850 nm. However, the addition of 700 nm SiNP (5H) makes a minor contribution to optical chirality in this wavelength range because its interaction with other meta-atoms is rarely affected by the switching of circular polarization. Similar phenomenon can also be observed in Fig. 5f where a location switching between the 400 nm AuNP in the middle and the 700 nm SiNP (5H) at the top flips the optical chirality above 850 nm (the region superimposed with yellow color in Fig. 5f), which reveals that the interaction in the 1 µm PS-400 nm AuNP-700 nm SiNP (5H) trimer structure makes a major contribution to the chiral features above 850 nm. The reconfigurability to assemble chiral meta-molecules into different isomers provides the possibility to bridge the molecular structures with the optical chirality to understand the origin of chirality at colloidal scale and to design the frequency-dependent chiral colloidal devices.

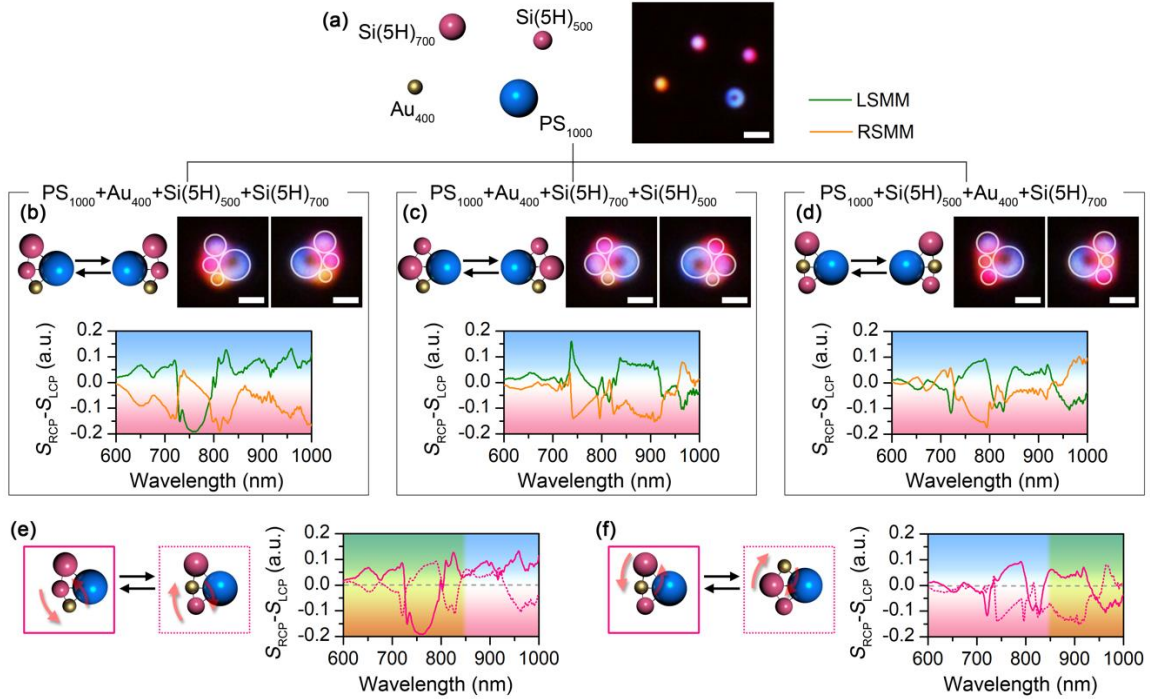

**FIGURE 5.** Reconfigurability of the experimentally assembled chiral meta-molecules. (a) Schematic and optical images of the dispersed meta-atoms. (b-d) Schematic, optical images, and differential scattering spectra of three sets of chiral meta-molecules composed of a 1 µm PS bead, a 400 nm AuNP, a 500 nm SiNP (5H), and a 700 nm SiNP (5H) with different configurations. (e-f) Comparison of the differential scattering spectra among the chiral meta-molecules. The solid and dashed curves correspond to the structures with solid and dashed outlines, respectively. In (b-d), the meta-molecules on the left are LHMMs, while the ones on the right are RHMMs. In (b-d), white circles are superimposed onto the dark-field optical images to guide the visualization of the molecular structures. In (e) and (f), the region superimposed with yellow shows mirror symmetry in the differential scattering spectra. Scale bars: 2 µm in (a); 1 µm in (b-d).

**Conclusions**

Mimicking atoms using colloids in a wide range of size and material, we demonstrate the atom-by-atom assembly of diverse chiral molecules at the colloidal scale, i.e., chiral meta-molecules. The invertible control of bonding/dis-bonding states between the meta-atoms is achieved by steering the light field, which allows us to reconfigure the molecular structures all-optically and tailor the optical chirality on-demand. With their structures observable under a microscope and the optical chirality several orders of magnitude stronger than the intrinsic chirality of organic molecules, the assembly of chiral meta-molecules opens a new opportunity to detect and investigate the origin of chirality at colloidal scale. Furthermore, the on-demand manufacture of meta-molecules represents a new platform to access metamaterials with tunable optical properties, which is important for multi-channel optical data processing and temporal response to complicated and unexpected external environment.

**Methods**

*Synthesis of SiNP*

Hydrogen-terminated amorphous SiNPs were synthesized following previously reported synthesis protocols [53]. In a nitrogen filled globe box, trisilane and n-hexane were added in a 10 mL titanium reactor. The reactor was then sealed, taken out from the glove box, and heated to the target temperature for ten minutes for complete decomposition of trisilane. The amount of trisilane determines the size of the SiNPs, while the amount of n-hexane determines the reaction pressure inside the reactor. The hydrogen concentration inside the SiNPs was controlled by changing the reaction temperatures. After the reaction, the reactor was cooled down to room temperature in an ice bath. The SiNPs were then washed with chloroform by centrifuging at 8000 rpm for 5 min. The precipitate was collected and dispersed in chloroform before use.

*Preparation of the thermoplasmonic substrate*

Cover slip with thickness of 120 μm was sonicated in isopropanol and deionized water for 15 minutes, respectively, to remove the contaminations. A 5 nm gold film was deposited onto the cover slip without any adhesion layer using thermal deposition below $1\times10^{-5}$ torr. The gold film was then annealed in an oven at 550 °C for two hours.

*Optical setup*

A 532 nm diode-pumped solid-state laser was used for optical heating. The laser was expanded with a 5× beam expander and directed to an inverted microscope (Nikon Ti-E). After going through a 100× oil-immersed objective (numerical aperture (NA): 0.5-1.3), the laser beam was focused on the thermoplamsonic substrate for optical heating. It should be noted that the smallest NA of 0.5 was used in the experiment to reduce the optical gradient force, as well as to improve the dark-field imaging. An oil-immersed condenser (NA:1.20-1.43) was used to focus the white light onto the samples for the dark-field imaging. Before the condenser, the polarization of the white light was controlled by a linear polarizer (LPNIRE100-B, Thorlabs Inc.) followed by a quarter-wave plate (AQWP10M-980, Thorlabs Inc.) to generate the circularly polarized light. The handedness of the circular polarization was further controlled by rotating the quarter-wave plate to tune the angle between the fast axis of the quarter-wave plate and the polarization axis of the linear polarizer. The dark-field scattering signal from the meta-atoms or meta-molecules was then directed to a spectrometer (Andor) for *in-situ* measurement while the heating laser beam was blocked by a notch filter.

*Optical assembly*

Before optical assembly, colloidal particles were dispersed into the CTAC solution at a concentration above 15 mM. A spacer of 120 µm in thickness was placed on the thermoplasmonic substrate. A drop of colloidal suspension was added into the chamber and confined by another cover slip on the top. A laser beam with the optical power of 0.1 mW was focused on the thermoplasmonic substrate to create the thermoelectric field for optical manipulation. Individual colloidal particles were picked up by the laser beam and assembled into the meta-molecules for *in-* *situ* spectroscopy. The on/off states of the laser beam were controlled by a motorized shutter (MFF102, Thorlabs Inc.) to achieve assembly/disassembly of the colloidal particles.

*Differential scattering spectroscopy*

After the assembly, scattering spectra from the chiral meta-molecules were taken at both left-handed and right-handed circular polarizations when the laser is on to hold the meta-molecules. Background spectra were taken when the laser is on without trapping of any particle at both left-handed and right-handed circular polarizations. After subtracting the background spectra, the scattering signal of the meta-molecules was normalized with the light source spectra. The intensity difference between the right-handed and left-handed scattering spectra was defined as the differential scattering spectra to characterize the optical chirality of the meta-molecules.

*Trapping stiffness*

The trapping stiffness of the 300 nm AuNPs was characterized by tracking the Brownian motion of the particles in the opto-thermoelectric trap. A monochromic charge-coupled device (CCD) camera (Andor) was used for the particle tracking at an exposure time of 10 milliseconds and a duration of 30 seconds. The position probability of the particles was fitted by the Gaussian function to obtain the variance of the Brownian motion $\sigma$ and the trapping stiffness was calculated by $k_T = 2k_B T/\sigma^2$.

*Numerical simulation*

To treat the optical response of spherical nanoparticles with the dielectric permittivity of $\varepsilon_p = n_p^2$ ($n_p$ is the refractive index of the nanoparticle) and a radius of $r_p$ located in the free space we have used the Mie scattering theory [54,55], which gives the following expression for normalized scattering cross section $Q_{sct}$:

$$Q_{sct} = \frac{2}{(kr_p)^2} \sum_{l=0}^{\infty} (2l+1)(|a_l|^2 + |b_l|^2) \qquad (1)$$

where $l$ defines the order of partial wave, $k$ is the wavenumber $\left(k = \frac{w}{c}n_h\right)$, $w$ is the angular frequency, $c$ is the speed of light in the vacuum, and $e_h = n_h^2$ is the dielectric permittivity of the surrounding medium. Electric and magnetic scattering amplitudes are given by $a_l = R_l^{(a)}/(R_l^{(a)} + iT_l^{(a)})$, $b_l = R_l^{(b)}/(R_l^{(b)} + iT_l^{(b)})$, and functions $R_l$ and $T_l$ can be expressed in the following form:

$$R_l^{(a)} = ny_l'(kr_p)y_l(nkr_p) - y_l(kr_p)y_l'(nkr_p),$$
$$T_l^{(a)} = nc_l'(kr_p)y_l(nkr_p) - c_l(kr_p)y_l'(nkr_p), \quad (2)$$

$$R_l^{(b)} = ny_l'(nkr_p)y_l(kr_p) - y_l(nkr_p)y_l'(kr_p),$$
$$T_l^{(b)} = nc_l(kr_p)y_l'(nkr_p) - c_l'(kr_p)y_l(nkr_p), \quad (3)$$

Here, $y_l(x) = \sqrt{\frac{\rho x}{2}} J_{l+1/2}(x)$, $c_l(x) = \sqrt{\frac{\rho x}{2}} N_{l+1/2}(x)$, $J_{l+1/2}(x)$ and $N_{l+1/2}(x)$ are the Bessel and Neumann functions, and prime means derivation.

FDTD simulations of the meta-molecules have been conducted by using CST Microwave Studio 2018. CST Microwave Studio is a full-wave 3D electromagnetic field solver based on finite-integral time domain solution technique. A nonuniform mesh was used to improve the accuracy near the nanoparticle boundary where the field concentration was significantly large and inhomogeneous.

The permittivity of gold was taken from ref [56], while the dielectric function of amorphous Si was taken from [57]. The permittivity of the liquid environment was set as 1.77. The trimer structures were excited by circularly polarized plane waves with the incident angle of 60 degrees. In order to simulate dark-field setup, the scattering spectra have been obtained by averaging of 18 forward scattering spectra calculated for different azimuthal angles from 0 to 360 degree with a step of 20 degrees. To calculate the electric and magnetic field distributions, we excited the trimers with a normally incident and circularly polarized plane wave. Such a simulation gives an idea of multipole excitation inside the meta-molecules.

In the model of Au chiral trimers, we used 330 nm (1), 260 nm (2), and 230 nm (3) AuNPs with the mutual gaps of 28 nm (1-2), 22 nm (1-3), and 60 nm (2-3). For the Si chiral trimmers, we used 700 nm (1), 500 nm (2), and 300 nm (3) SiNPs with the mutual gaps of 20 nm (1-2), 17 nm (1-3), and 14 nm (2-3). The gaps between nanoparticles were fitted in order to get a good agreement between numerical and experimental results.

**Acknowledgements**

L.L., X.P. and Y.Z. acknowledge the financial supports of the National Science Foundation (NSF-CMMI-1761743), the Army Research Office (W911NF-17-1-0561), the National Aeronautics and Space Administration Early Career Faculty Award (80NSSC17K0520), and the National Institute of General Medical Sciences of the National Institutes of Health (DP2GM128446). B.A.K. and T.J. acknowledge the financial support of this work from the Robert A. Welch Foundation (grant no. F-1464) and the National Science Foundation through the Center for Dynamics and Control of Materials: an NSF MRSEC under Cooperative Agreement No. DMR-1720595. We also thank the Texas Advanced Computing Centre at The University of Texas at Austin for providing HPC resources that have contributed to the research results reported within this paper. URL: http://www.tacc.utexas.edu.

**Author contributions**

L.L. and Y.Z. conceived the idea. L.L worked on the experiments and analysed the data. A.K., S.L., and A.A. conducted the numerical simulations. T.J. and B.A.K. synthesized the SiNPs. Y.Z. supervised the project. L.L. and Y.Z. wrote the manuscript. All authors participated in the discussion of the results and proofread the manuscript.

**Competing financial interests**

The authors declare no competing financial interests.

**Appendix A. Supplementary data**

Supplementary data associated with this article can be found, in the online version.

**Data availability**

The data that support the plots within this paper and other findings of this study are available from the corresponding author upon reasonable request.

**Supplementary Video**

**Supplementary video 1.** Assembly process of both left-handed and right-handed chiral meta-molecules composed of a 1 μm PS, a 300 nm AuNP, a 300 nm SiNP (5H) and a 500 nm SiNP (5H) at 2× speed.

**Supplementary Figures**

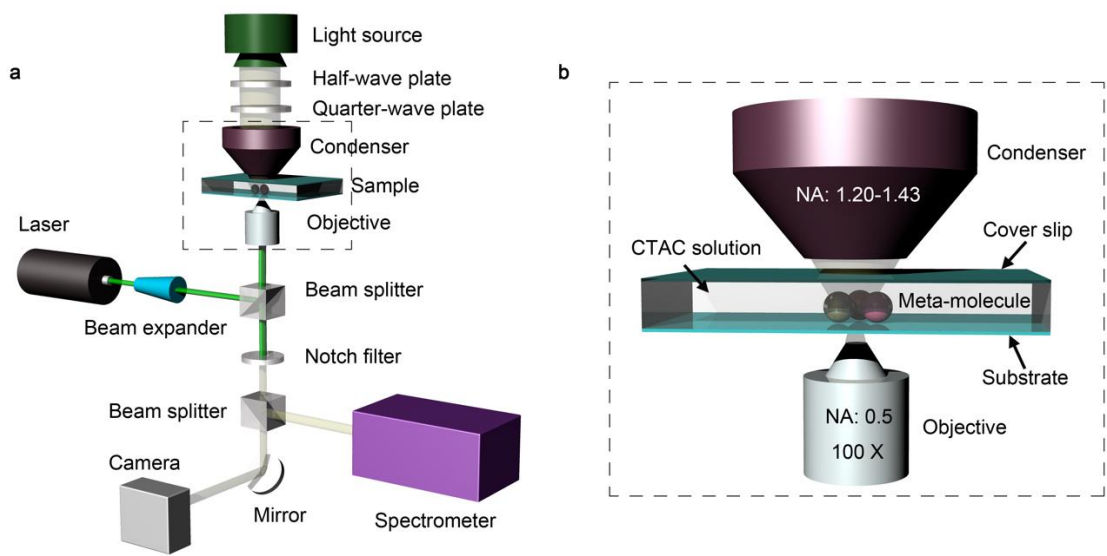

**Supplementary Figure 1.** (a) The optical setup for optical assembly and in-situ measurement. (b) Detailed setup for the dark-field scattering measurement.

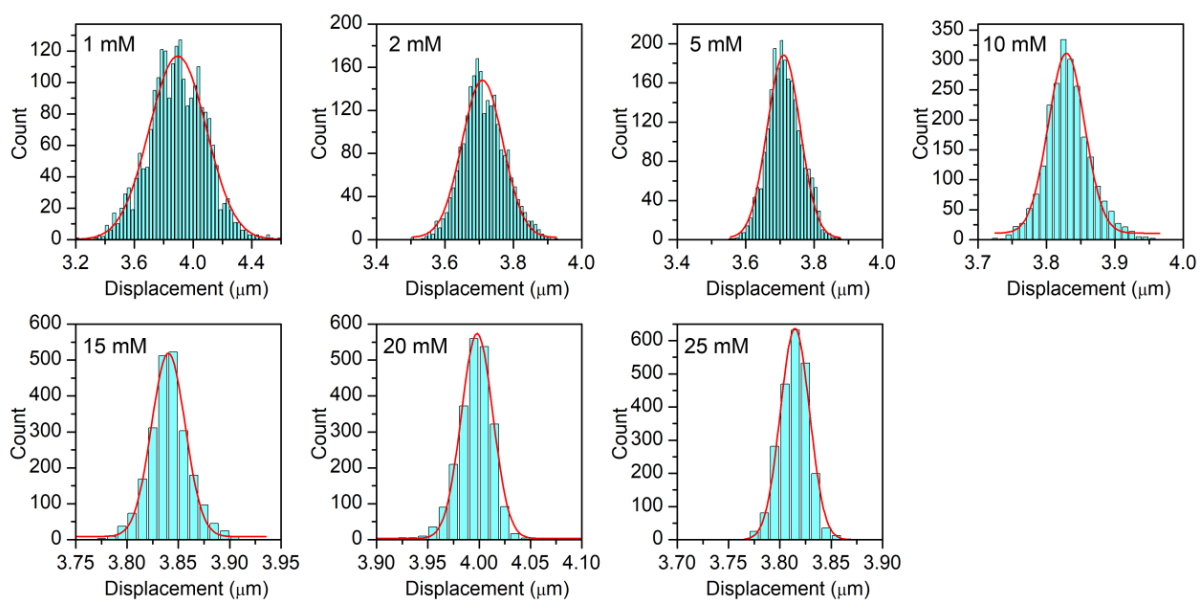

**Supplementary Figure 2.** Representative measurements of positional displacement of single 300 nm AuNPs trapped in the thermoelectric field at different CTAC concentrations.

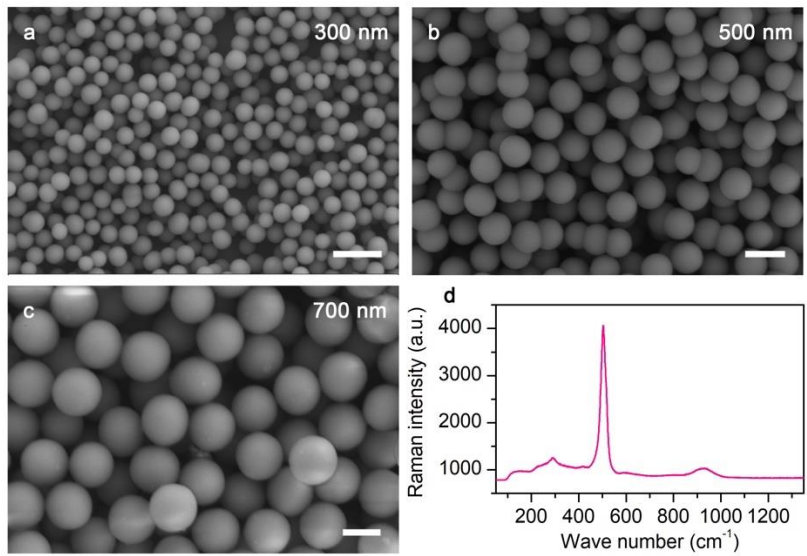

**Supplementary Figure 3.** a-c, Scanning electron microscopy (SEM) images of the silicon nanoparticles (SiNPs) with different sizes: a, 300 nm; b, 500 nm; c, 700 nm. d, Raman spectrum of 300 nm SiNP (5H). Scale bars: 1 μm (a); 600 nm (b); 500 nm (c).

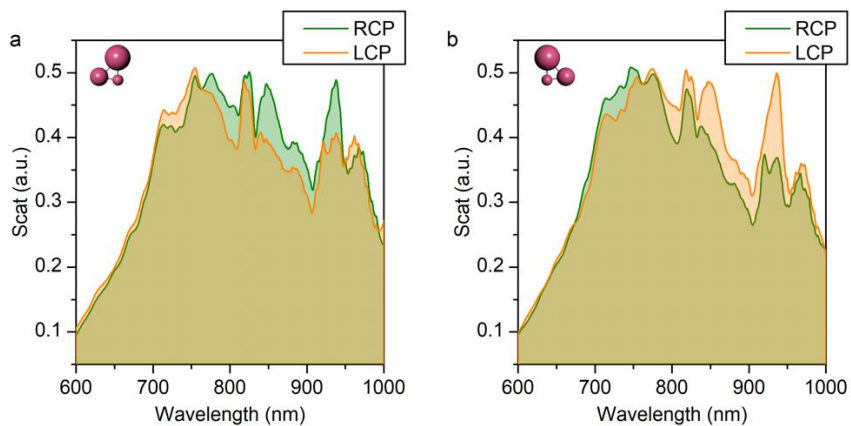

**Supplementary Figure 4.** Scattering spectra of both the left-handed and right-handed Si trimers under different circularly polarized light. a, left-handed Si trimer; b, right-handed Si trimer.

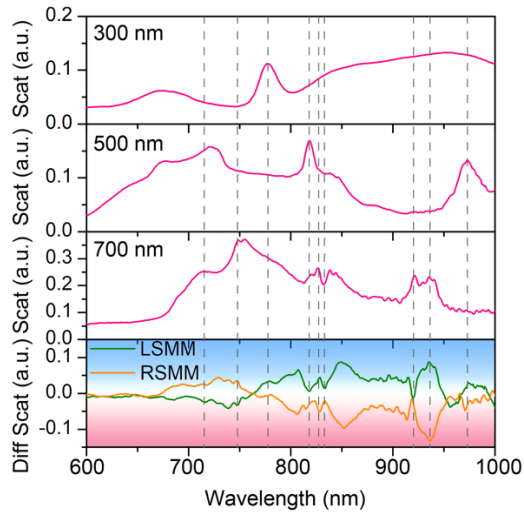

**Supplementary Figure 5.** Single nanoparticle spectra and the differential scattering spectra of the Si trimer. The dash lines show matching between the single nanoparticle spectra and the differential scattering spectra. The sizes of the silicon nanoparticles in the dimers are 300 nm, 500 nm, and 700 nm, respectively, which are labelled at the top left corners.

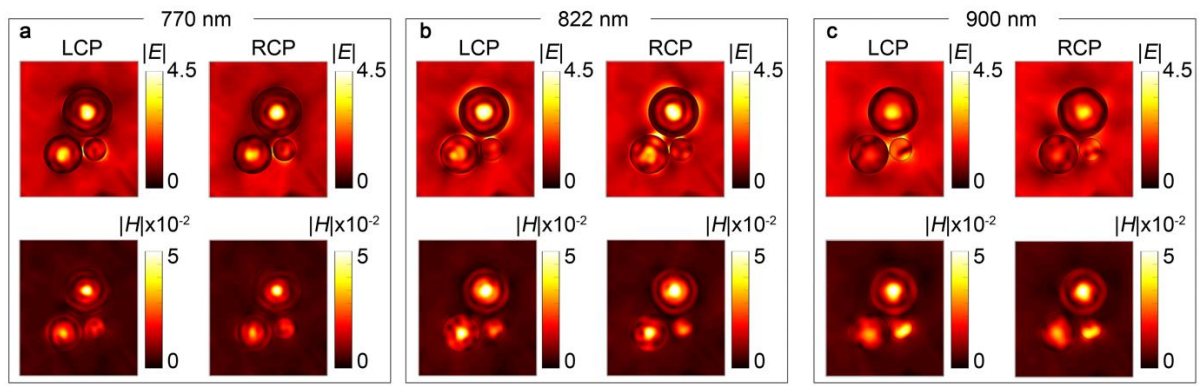

**Supplementary Figure 6.** Electric and magnetic field distribution of the Si chiral trimer under different circular polarizations. a, 770 nm; b, 822 nm; c, 900 nm.

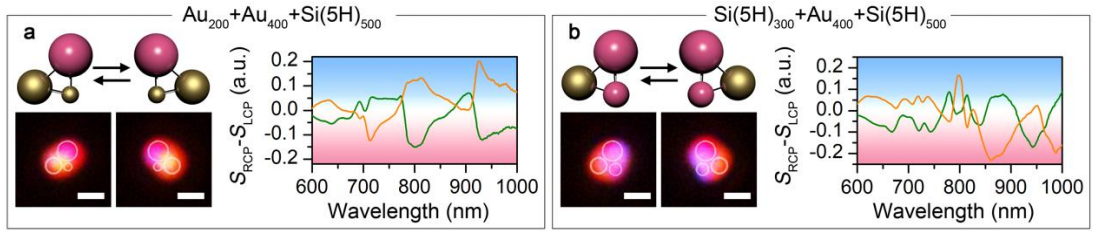

**Supplementary Figure 7.** Hybrid chiral meta-molecules. Schematic, optical images, and differential scattering spectra of hybrid chiral trimers composed of a, a 200 nm AuNP, a 400 nm AuNP, and a 500 nm SiNP (5H); b, a 300 nm SiNP (5H), a 400 nm AuNP, and a 500 nm SiNP (5H). Scale bar: 1 µm.

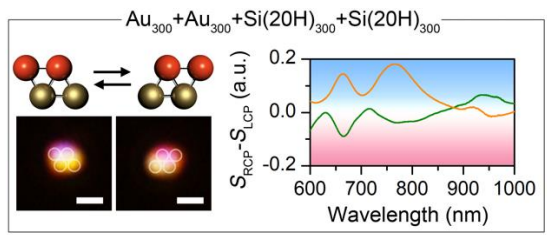

**Supplementary Figure 8.** Schematic, optical images, and differential scattering spectra of C-chiral tetramer composed of two 300 nm AuNPs and two 300 nm SiNPs (20H). Scale bar: 1 µm.

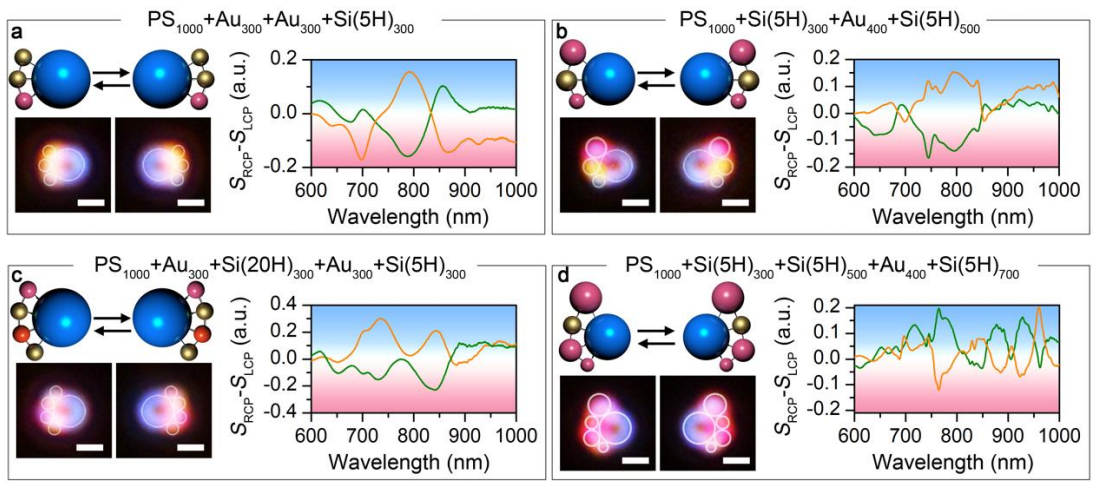

**Supplementary Figure 9.** Saturn-ring chiral meta-molecules. Schematic, optical images, and differential scattering spectra of the Saturn-ring chiral meta-molecules composed of a, a 1 µm PS

bead, two 300 nm AuNPs, and a 300 nm SiNP (5H); b, a 1 µm PS bead, a 300 nm SiNP (5H), a 400 nm AuNP, and a 500 nm SiNP (5H); c, a 1 µm PS bead, two 300 nm AuNPs, a 300 nm SiNP (20H) and a 300 nm SiNP (5H); d, a 1 µm PS bead, a 300 nm SiNP (5H) a 500 nm SiNP (5H), a 400 nm AuNP, and a 700 nm SiNP (5H). Scale bars: 1 µm.

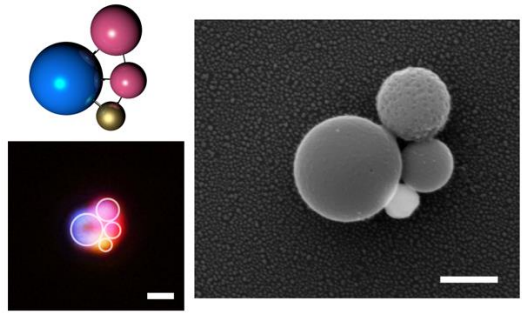

**Supplementary Figure 10.** Schematic, optical image, and SEM image of a chiral meta-molecule composed of a 1 µm PS, a 400 nm AuNP, a 500 nm SiNP, and a 700 nm SiNP. Scale bars: 1 µm (optical image); 500 nm (SEM).

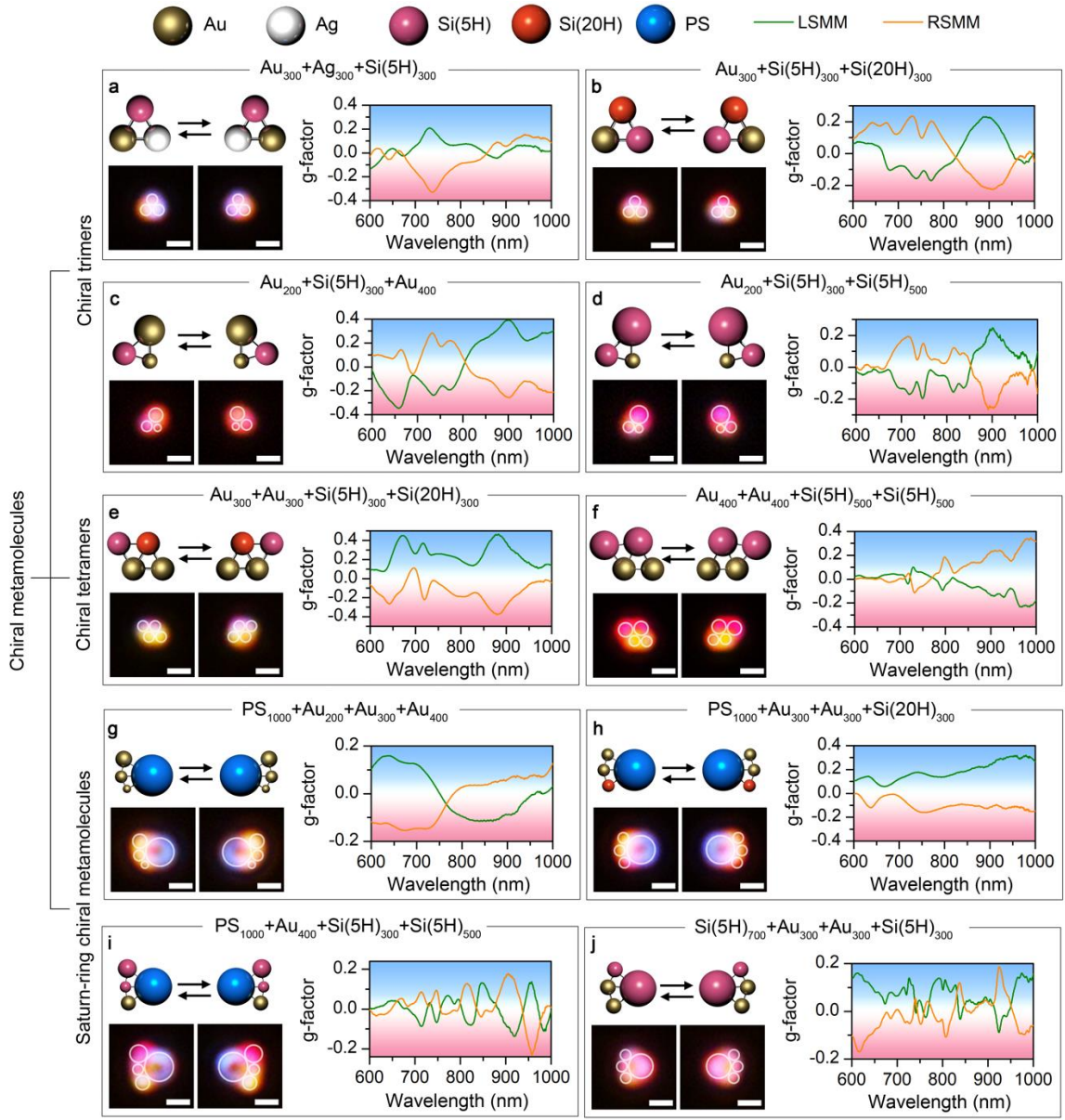

**Supplementary Figure 11.** G-factor of the chiral meta-molecules. Schematic, optical images and g-factor of chiral trimers composed of (a) an AuNP, an AgNP and a SiNP (5H) with diameter of 300 nm; (b) an AuNP, a SiNP (5H) and a SiNP (20H) with diameter of 300 nm; (c) a 200 nm AuNP, a 300 nm SiNP (5H), and a 400 nm AuNP; (d) a 200 nm AuNP, a 300 nm SiNP (5H) and a 500 nm SiNP (5H). Schematic, optical images and g-factor of chiral tetramers composed of (e) two 300 nm AuNP, a 300 nm SiNP (5H), and a 300 nm SiNP (20H); (f) two 400 nm AuNPs and two 500 nm

SiNPs (5H). Schematic, optical images g-factor of Saturn-ring chiral meta-molecules composed of (g) a 1 μm PS bead, a 200 nm AuNP, a 300 nm AuNP, and a 400 nm AuNP; (h) a 1 μm PS bead, two 300 nm AuNPs, and a 300 nm SiNP (20H); (i) a 1 μm PS bead, a 400 nm AuNP, a 300 nm SiNP (5H), and a 500 nm SiNP (5H); (j) a 700 nm SiNP (5H), two 300 nm AuNPs, and a 300 nm SiNP (5H). The meta-molecules on the left are LHMMs, while the ones on the right are RHMMs. Scale bar: 1 μm.

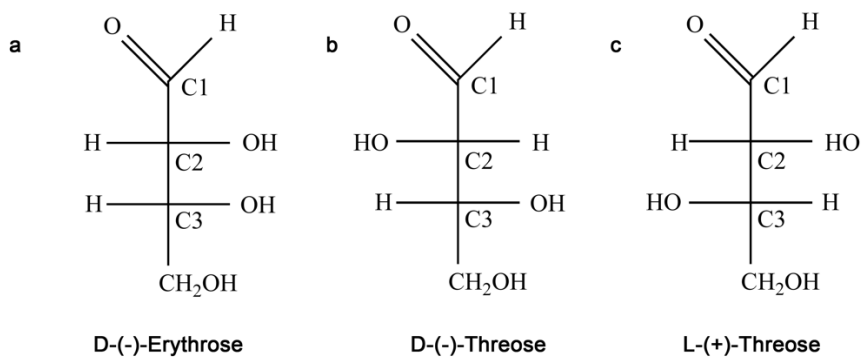

**Supplementary Figure 12.** Molecular structures of erythrose and threose. a, D-erythrose shows negative optical rotation; b, D-threose shows negative optical rotation; c, L-threose shows positive optical rotation. D-threose and L-threose are enantiomers, while threose and erythrose are diastereomers.

# Supplementary Note

**Calculation of the interparticle interaction potential**

A full consideration of the electrostatic interaction $U_e$, Van der Waals interaction $U_{vdw}$, and depletion interaction $U_d$ is taken here to calculate the interaction potential between two 300 nm AuNPs:

$$U_{total} = U_e + U_{vdw} + U_d \qquad (1)$$

Electrostatic interaction

Considering the CTAC adsorption on the AuNPs, the AuNPs are treated as positively charged spheres. The electrostatic interaction potential between two 300 nm AuNPs is written as [1,2]:

$$U_e = 32\pi \varepsilon_s r_p \left(\frac{k_B T}{ze}\right)^2 \left[\tanh\left(\frac{ze y_p}{4 k_B T}\right)\right]^2 \exp(-\kappa r) \qquad (2)$$

where $\varepsilon_s$ is the solvent permittivity ($7.08 \times 10^{-10}$ F/m for water at 20 °C), $r_p$ is particle radius (150 nm), $k_B$ is the Boltzmann constant, $T$ is the absolute temperature, $z$ is the particle charge valence, $e$ is the elemental charge, $y_p$ is the zeta potential of the particle, $\kappa$ is the inverse Debye length which is a function of CTAC concentration, and $r$ is the gap size between two particle surface. The calculated $U_e$ at different CTAC concentration is summarized in Supplementary Fig. 13.

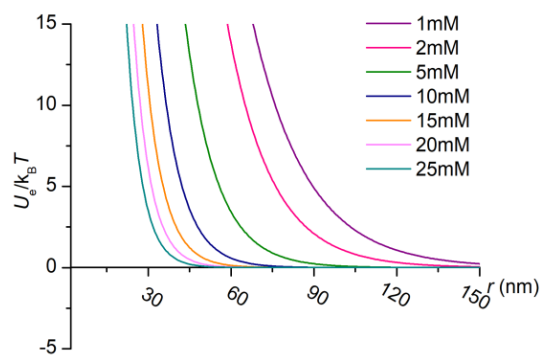

**Supplementary Figure 13. Electrostatic interaction potential between two 300 nm AuNPs as a function of the inter-particle gap at different CTAC concentrations.**

Van der Waals interaction

The Van der Waals interaction potential is given by

$$U_{vdw} = -\frac{Ar_p}{12r} \qquad (3)$$

where $A$ is the retarded Hamaker constant of Au [3], which is plotted in Supplementary Fig. 14a. The calculated $U_{vdw}$ as a function of the inter-particle gap is summarized in Supplementary Fig. 14b.

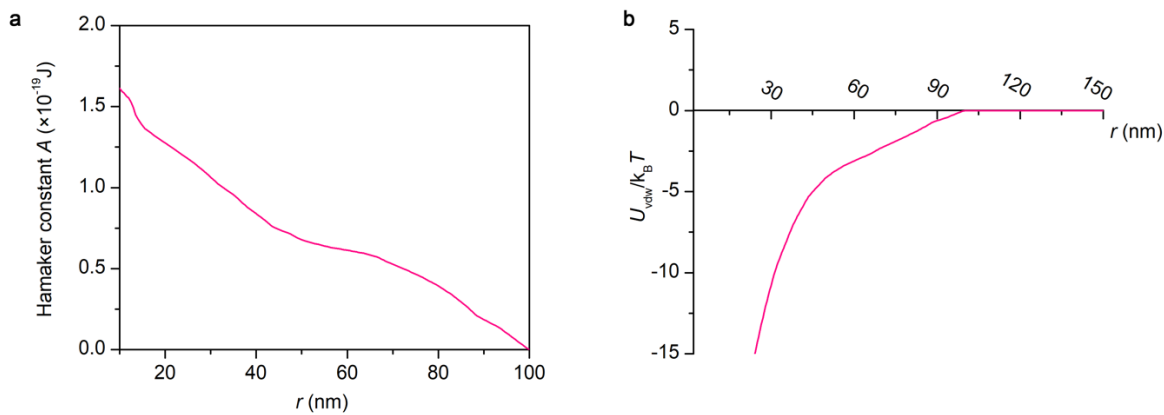

**Supplementary Figure 14. Calculated Van der Waals interaction potential. a,** Retarded Hamaker constant of AuNPs as a function of the inter-particle gap. **b,** Van der Waals interaction potential between two 300 nm AuNPs as a function of the inter-particle gap.

Depletion attraction

The thermophoretic migration of CTAC micelles leads to the depletion of CTAC micelles (known as depletants) at the inter-particle gap, which give rise to an osmotic pressure on the particles. The depletion attraction potential is given by

$$U_d = -\Delta V \Delta P \tag{4}$$

where $\Delta P$ is the osmotic-pressure difference, while $\Delta V$ is the depletion volume of the CTAC micelles. It should be noted that CTAC micelles are highly charged depletants and the electrostatic interaction should be considered to calculate $\Delta V$ [1,2]:

$$\Delta V = \rho[\frac{4}{3}(r_p + r_d^{eff})^3(1 - \frac{3}{4}(r_p + r_d^{eff} + r)(r_p + r_d^{eff})^{-1} + \frac{1}{16}(r_p + r_d^{eff} + r)^3(r_p + r_d^{eff})^{-3})] \tag{5}$$

where $r_d^{eff}$ is the effective radius of the CTAC micelles where the micelle-particle interaction potential is taken into account:

$$r_d^{eff} = r_d + m\kappa^{-1} \tag{6}$$

To evaluate the value of $r_d^{eff}$, we calculate the electrostatic interaction potential between a 300 nm AuNP and a CTAC micelle at by [1,2]

$$U_e(dp) = 4\pi\varepsilon_s y_d y_p \frac{r_d r_p}{(r_d + r_p + r_{dp})}\exp(-\kappa r_{dp}) \tag{7}$$

where $y_d$ is the surface potential of the CTAC micelles, $r_d$ is the hard-sphere radius of CTAC micelles (~3 nm), and $r_{dp}$ is the surface-to-surface distance between the micelle and the AuNS. The micelle-particle interaction potential as a function of the micelle-particle gap at 10 mM is plotted in Supplementary Fig. 15a. We effective micelle-particle gap is estimated at $U_e(dp) = 1k_BT$, giving $m=3.99$. We further calculate the $r_d^{eff}$ value at different CTAC concentration, which is summarized in Supplementary Table 1.

**Supplementary Table 1. Effective radius of the CTAC micelles at different CTAC concentration**

| CTAC concentration (mM) | 1 | 2 | 5 | 10 | 15 | 20 | 25 |
|---|---|---|---|---|---|---|---|
| $r_d^{eff}$ (nm) | 81.5 | 66.6 | 47.6 | 35.8 | 30.2 | 26.7 | 24.3 |

It should also be noted that micelles is a kind of soft depletant compared with other hard-sphere depletants, and the osmotic-pressure difference $\Delta P$ is given by

$$\Delta P = bnk_BT \tag{8}$$

where $n$ is the number density of the CTAC micelles, while $b$ is the depletion fraction. A fitting between experimental data with the theoretical model gives a value of $b=20\%$. $n$ is calculated by

$$n = \frac{N_A}{N_{agg}}(c_s - c_{cmc}) \tag{9}$$

where $N_A$ is the Avogadro constant, $N_{agg}$ is the aggregate number of the CTAC micelles which is a function of the CTAC concentration $c_s$ [4], and $c_{cmc}$ is the critical micelle concentration of CTAC (~0.13 mM). Finally, the depletion attraction potential at different CTAC concentrations is calculated and summarized in Supplementary Fig. 15b.

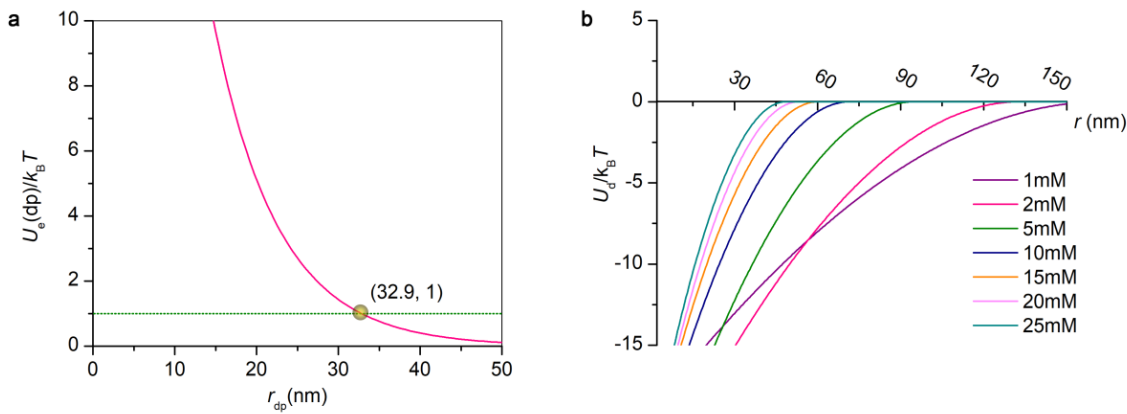

**Supplementary Figure 15. Calculated depletion interaction potential. a,** Electrostatic interaction potential between a 300 nm AuNP and a CTAC micelles as a function of the micelle-particle gap $r_{dp}$ at 10 mM. The green dot line shows $U_e(dp) = 1k_BT$. The intersection point is indicated as the yellow ball. **b,** Depletion interaction potential between two 300 nm AuNPs as a function of the inter-particle gap at different CTAC concentrations.